\documentclass[rmp,aps,floatfix,showkeys,12pt]{revtex4}
\usepackage[]{graphicx}
\usepackage{amsmath}
\usepackage{amssymb}
\usepackage{times}
\usepackage{natbib}
\usepackage{color}
\begin{document}

\title{Ecological-economic modelling of interactions between wild and commercial bees and pesticide use}

\author{Adam Kleczkowski}
\email{ak@maths.stir.ac.uk}
\affiliation{Computing Science and Mathematics, School of Natural Sciences, University of Stirling, UK}

\author{Ciaran Ellis}
\affiliation{Biological and Environmental Studies, School of Natural Sciences, University of Stirling, UK}

\author{Dave Goulson}
\affiliation{School of Life Sciences, University of Sussex, UK}

\author{Nick Hanley}
\affiliation{Dept.\ of Geography and Sustainable Development, University of St Andrews, UK}

\begin{abstract}
The decline in extent of wild pollinators in recent years has been partly associated with changing farm practices and in particular with increasing pesticide use. In this paper we combine ecological modelling with economic analysis of a single farm output under the assumption that both pollination and pest control are essential inputs. We show that the drive to increase farm output can lead to a local decline in the wild bee population. Commercial bees are often considered an alternative to wild pollinators, but we show that their introduction can lead to further decline and finally local extinction of wild bees. The transitions between different outcomes are characterised by threshold behaviour and are potentially difficult to predict and detect in advance. Small changes in economic parameters (input prices) and ecological parameters (wild bees carrying capacity and effect of pesticides on bees) can move the economic-ecological system beyond the extinction threshold. We also show that increasing the pesticide price or decreasing the commercial bee price might lead to re-establishment of wild bees following their local extinction. Thus, we demonstrate the importance of combining ecological modelling with economics to study the provision of ecosystem services and to inform sustainable management of ecosystem service providers. 
\end{abstract}


\maketitle

\newtheorem{theorem}{Proposition}


\section{Introduction}

\label{sec:intro} Globally, 35\% of food crops are at least partly dependent on insect pollination \citep{Klein2007}.  Ensuring sufficient pollination of these crops will be challenging in the future; the fraction of agriculture made up by insect-pollinated crops is increasing \citep{Aizen2009}, while wild pollinator populations are threatened by both habitat loss \citep{Winfree2009} and agricultural intensification, which are thought to be the main causes of reported declines in diversity in the EU and in the USA \citep{Biesmeijer2006,Cameron2011}.  

For some crops, honeybees are used to supplement or substitute wild pollinators, along with other commercial pollinators such as laboratory bred bumblebees \break \citep{Velthuis2006}. While commercial pollinators are often assumed to be adequate substitutes for wild pollinators (though see \citep{Brittain2013,Hoehn2008}), the use of commercial pollinators is itself not without risk.  Honeybees have suffered losses in recent years due to the abandonment of hives (Colony Collapse Disorder) and the \emph{Varroa} mite \citep{Cox-Foster2007}.  Relying on commercial pollinators such as honeybees puts farmers at risk from shocks of this kind, with consequent implications for farm profits over time.

Given the risks around the supply of pollination services from commercial bees, maintaining viable wild pollinator populations is likely to be crucial to sustaining the production of insect-pollinated crops into the future \citep{Winfree2007}.  The potential costs of the loss of local pollination services is illustrated by the need for pollination using farm workers in Sichuan, China, following the loss of local insect pollinator populations \citep{Partap2001}.  Whilst this was a viable
option when wages were cheap, a 10--fold rise in wages over the last 10 years has led to the abandonment of apple production \citep{Partap2012}.  One of the factors implicated in this local extinction and in declines elsewhere, is the use of pesticides, or specifically, insecticides.  There is growing evidence of negative effects of realistic levels of commonly used insecticides on population determining traits such as reproductive rates, foraging rates and navigation in bees  \citep{Henry2012,Mommaerts2010,Whitehorn2012,Goulson2013}.  Awareness of this evidence has led to the temporary banning of a very widely used group of insecticides -- neonicotinoids -- within the European Union.

Farmers who grow insect-pollination dependent crops face a trade-off in their use of pesticides, since whilst this reduces crop damages, it also has potential negative effects on the local supply of pollination services from both wild and commercial pollinators. This is the issue that we study in this paper. To investigate the links between commercial pollinator use, pesticide use and wild pollinator populations we 
present an ecological-economic model which links crop yields to pollinator
numbers and pesticide use, and study how the optimal strategy
depends on the level of farm output and other parameters. 

There are a number of key results in this paper:  Firstly,
by introducing commercial bees the farmer can achieve higher target output
values than by relying on wild pollinators alone for the same level of cost. Thus, it is rational to substitute commercial for wild bees and to increase the use of insecticides.  Secondly, the resulting increase in
pesticide use buffered by usage of commercial bees may lead to a severe reduction in wild pollinator population, even without direct competition between different bee
populations. Thirdly, under certain economic conditions, the system becomes unstable and small changes in prices or environmental conditions can result in the local extinction of wild pollinators. As in this paper we are modelling a single farm only, all references to extinction mean that \textit{locally} the wild bee population will tend to zero.

Finally, we discuss a number of options that are available to restore a wild
bee population even if it is currently \textit{locally} extinct. These include an increase in the pesticide price (or application of a pesticide tax), increase in the carrying capacity characterising the wild bee populations, or a change in the type of insecticides used on the farm. Perhaps surprising, a decrease in the price of commercial bees can under some conditions help the wild bee population indirectly by allowing farmers to reduce their reliance on pesticides.

\section{Modelling Framework}

\label{sec:methods}

The model describes a single farm and its
surrounding ecosystem supporting wild bees. 
Farmer returns consist of two components, output (yield) and costs. Output
is assumed to follow a Cobb-Douglas production function with two inputs:
pollination services and pest control. Wild bees together with commercial bees (if present) provide pollination services for the agricultural production on the farm, whereas pest control is achieved by application of pesticides. 

\subsection{Output}
We identify the following key economic and ecological factors that need to be included in the model: (i) Both pollination and pest control are essential inputs; the output
is zero if \textit{either} of the inputs is zero. (ii) Increasing pesticide and pollination inputs increases  production, but only up to a certain limit. (iii) Wild and commercial bees are substitutes, but they can perform differently for different crops. (iv) Pesticides affect the population of both wild and commercial bees, although the farmer can balance the effect on commercial bees by increasing their supply. (v) Commercial bees, when they are present at the farm, are assumed not to compete directly with wild bees in terms of foraging for food.%
\footnote{%
There is mixed evidence on this relationship in the scientific literature
\citep{Garibaldi2011}.}

Thus, we assume a general form for the production term

\begin{equation}
q(x_{1},x_{2})= A \left(B(y,z,x_1)\right)^\alpha \left(P(x_{1})\right)^\beta
,  \label{eq:production}
\end{equation}%

where $B$ is a function representing pollination services and $P$ is a function representing pest control. The level of pesticide application is denoted by $x_1$, the potential for pollination by wild bees is denoted by $y$, and pollination by commercial bees is denoted by $z$. $\alpha $ is the output elasticity of pollination services and $ \beta $ is the output elasticity of pest control. Further, $A$ represents total
factor productivity, and without loss of generality we assume $A=1$. Functions $B(y,z,x_1)$ and $P(x_1)$ are assumed to be zero if inputs are zero, $B(0,0,x_1)=0$ (for any value of $x_1$) and $P(0)=0$. 

The pollination services are assumed to increase if the actual population of wild bees, $y$, or actual population of commercial bees, $z$, increase. The second derivatives can be either zero (linear dependence, as in the standard Cobb-Douglas model) or negative (representing decreasing capacity of bees to pollinate); in the latter case we will assume that $B$ reaches an asymptote for large values of $y$ and $z$. In order to model the effect of pesticides on pollination, we assume that $B(y,z,x_1)$ is a decreasing function of $x_1$. As there is not enough experimental evidence to suggest a precise dependence, we assume a linear decrease for both the wild bees and commercial bees, so that

\begin{equation}
\begin{array}{l*{10}{c}c}
y&=&\kappa-g_1 x_1 &\text{~~if~~} \kappa-g_1 x_1>0 &\text{~~and 0 otherwise}\\[10pt]
z&=&x_2-g_2 x_1 &\text{~~if~~} x_2-g_2 x_1>0 &\text{~~and 0 otherwise,}
\end{array}
\label{eq:bees}
\end{equation}%

where $\kappa$ is the carrying capacity of the \textit{local} wild bee population and $g_1$, $g_2$ represent effect of pesticides on bee population. $x_2$ represents the number of commercial bees as introduced by the farmer and needs to be distinguished from $z$ which is the actual size of the population taking into account death due to pesticides. Thus, the farmer buys $x_2$ bees, but only $z$ survive and contribute to pollination. $x_2$ together with $x_1$ are two control variables in the model.

As wild and commercial bees are substitutes, we further simplify the function $B(y,z,x_1)$ to depend on a linear combination of $y$ and $z$, $y+u z$ rather than on these two variables separately. $u$ represents a relative efficiency of commercial bees compared to wild bees. 

Finally, we assume that $P(x_1)$ is an increasing function of $x_1$ (positive first derivative) but with either a zero second derivative (standard Cobb-Douglas model) or reaching an asymptote for large values of $x_1$.

\subsection{Cost}

Wild pollination services come free (unless costly actions are taken to improve habitat), whereas other costs are assumed to be
linear in output. The cost of pesticides is $w_1$ per unit and the cost of
commercial bees is $w_2$ per unit. We assume that the farmer minimises the
cost function:

\begin{equation}
c\left( {x_{1},x_{2}}\right) =w_{1}x_{1}+w_{2}x_{2}.  \label{eq:costbase}
\end{equation}%
The minimisation is subject to satisfying a total output constraint

\begin{equation}
q\left( {x_{1},x_{2}}\right) = \bar{q},  \label{eq:profitbase}
\end{equation}%

where $\bar{q}$ is the exogenous target output. Given $\bar{q}$, the levels
of pesticides, $x_{1}$, and commercial bees, $x_{2}$, are chosen by the
farmer based upon cost minimisation.

\section{Results}
\label{sec:results}

\subsection{Simplified model}
We initially consider a simplified model in which functions $B$ and $P$ are assumed to be linear (i.e the standard Cobb-Douglas formulation). We also assume constant returns to scale and hence $\beta=1-\alpha$. Finally, we neglect the effect of pesticides on commercial bees, hence $g_2=0$, and assume that wild and commercial bees are perfect substitutes (so $u=1$). We will relax these simplifying assumptions later in the paper. Thus,

\begin{equation}
q\left( {x_{1},x_{2}}\right) ={\left( {\kappa -g_1 x_{1}+x_{2}}\right)
^{\alpha }}{x_{1}^{1-\alpha }}= \bar{q}\,,  \label{eq:profit1}
\end{equation}%
whereas for $\kappa-g_1 x_1\leq 0$,

\begin{equation}
q\left( {x_{1},x_{2}}\right) ={x_{2}^{\alpha }}{x_{1}^{1-\alpha }}= \bar{%
q}.  \label{eq:profit1for0}
\end{equation}%

The farmer's decision problem can therefore be written as%

\begin{equation}
\label{eq:problem}
\min_{\mbox{\large ${x_1\geq 0,\, x_2\geq 0,\quad y=\kappa-g_1 x_1\geq 0} \atop {q \left( {x_1,x_2} \right)}= \bar{q}$}} c \left( {x_1,x_2} \right) ,
\end{equation}

\subsubsection{Optimisation}

\begin{figure}[htbp]
\begin{tabular}{cc}
\includegraphics[angle=0, width=6cm]{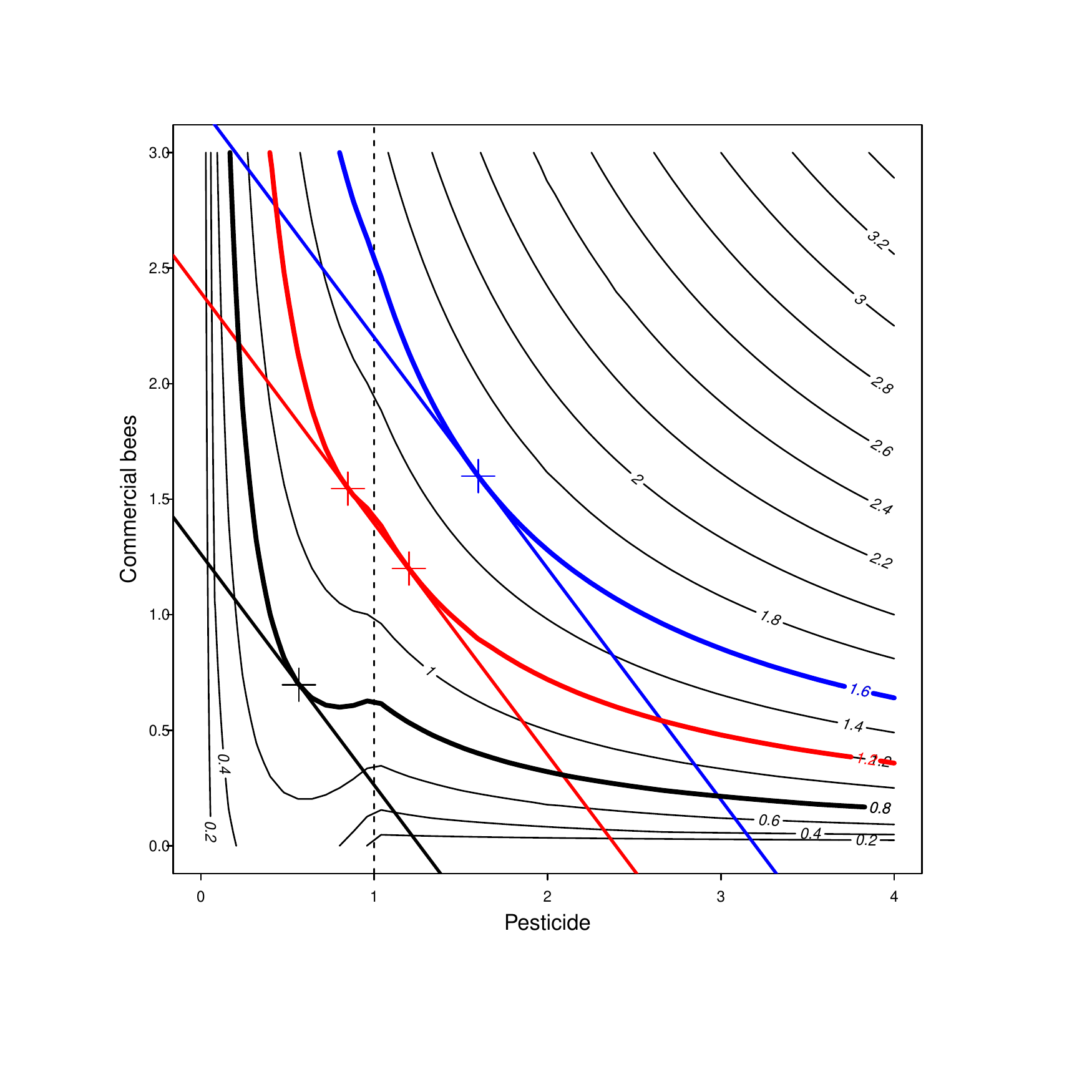} & %
\includegraphics[angle=0, width=6cm]{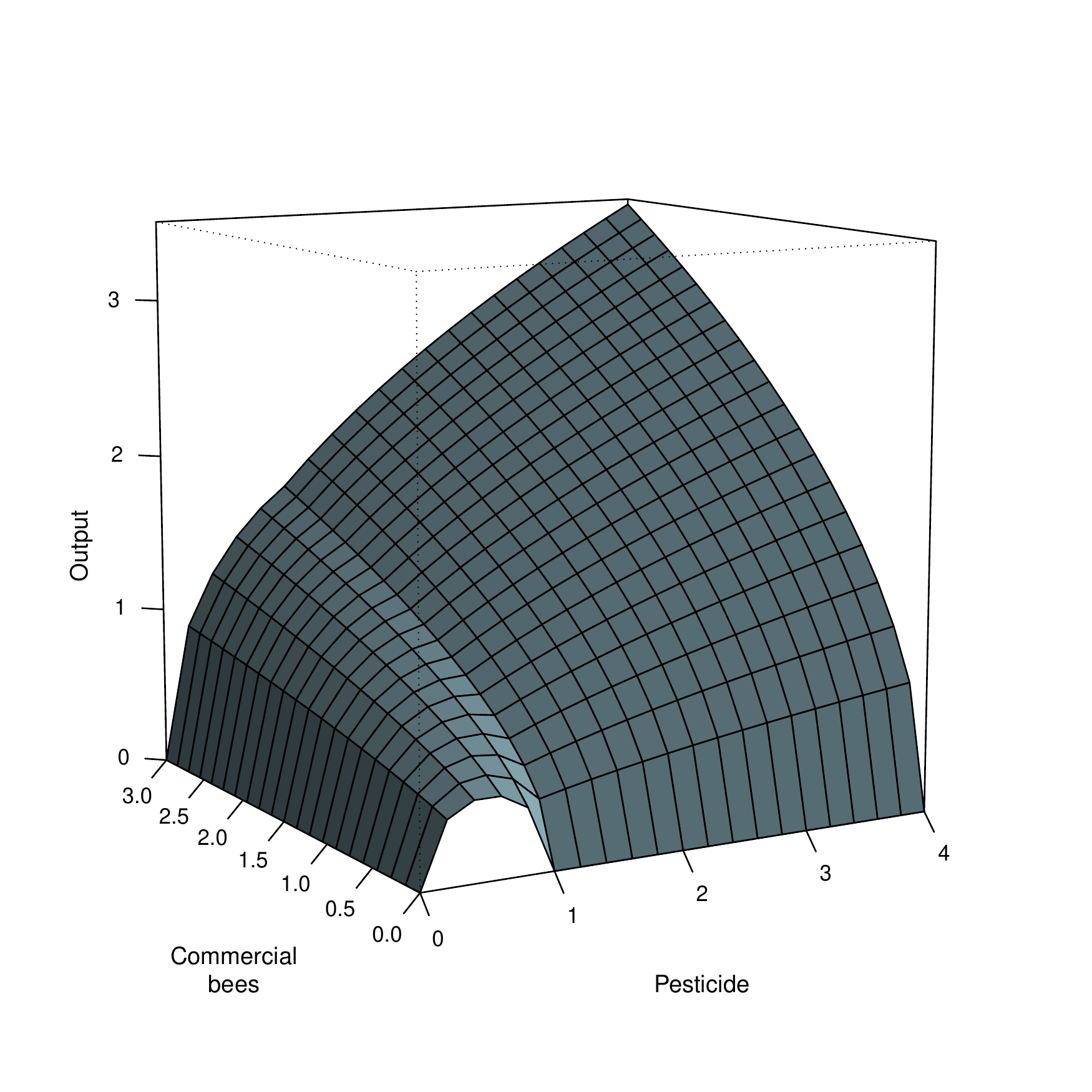} \\ 
& 
\end{tabular}%
\caption{Farm output as a function of $x_1$ and $x_2$, with values on
contour lines giving $q(x_1,x_2)$. Straight lines correspond to $w_1 x_1+w_2
x_2=\mbox{const}$. Isoquant lines and the corresponding minimum cost lines are shown
for $\bar{q}=0.8$ (black), $\bar{q}=1.2$ (red) and $\bar{q}=1.6$ (blue).
Vertical line corresponds to $y=\protect\kappa-gx_1=0$ and wild bees are
extinct to the right of it. On the right, output is shown in a perspective
plot. Other parameters: $\protect\kappa=1$, $g=1$, $w_1=1$, $w_2=1$ and $%
\protect\alpha=1/2$.\vskip2cm}
\label{fig:fig1}
\end{figure}

We assume that the farmer wants to achieve a target level of output, $\bar{%
q}$, and hence the constraint becomes $q\left( x_{1},x_{2}\right) =\bar{q}$ (this situation can arise if farmer is contracted to achieve a certain level of production).
A pair $\left( \hat{x}_{1},\hat{x}_{2}\right) $ represents a management
strategy that she can choose to achieve this. The minimisation process can
be represented on the $\left( x_{1},x_{2}\right) $ plane, with the isoquants 
$q\left( x_{1},x_{2}\right) =\bar{q}$ representing the output constraint,
Fig. \ref{fig:fig1}. In the simple case when there is no interaction between
pesticides and commercial bees, Eqn. (\ref{eq:costbase}), the cost function
is represented by a straight line with a negative slope, $-w_{1}/w_{2}$ and
the intercept $c/w_{2}$, see Fig. \ref{fig:fig1} with $c$ representing the cost. The procedure for optimizing, Eqn. (\ref{eq:problem}), can then be interpreted as finding a minimal value of $c$
such that the straight line still crosses the isoquant line corresponding to
the given value of $\bar{q}$, see Fig. \ref{fig:fig1}. The optimum value of $%
c$ corresponds to the straight line that is tangent to the isoquant line; if
more than one such line exists (representing local minima), the one with the
smallest value of $C$ is chosen. Fig. \ref{fig:fig1} gives three examples of
such lines for different values of $\bar{q}$.

As the cost is linear in $x_{1}$ and $x_{2}$, following standard
microeconomic theory \citep{GravelleRees2004}, this procedure can be recast
in terms of the following two conditions that can be used to find the
conditional factor demands $\hat{x}_{1}$ and $\hat{x}_{2}$ that minimise the
costs of producing $\bar{q}$ units of output:

\begin{subequations}
\begin{eqnarray}
\displaystyle\frac{\frac{\partial q(x_{1},x_{2})}{\partial x_{1}}}{\frac{%
\partial q(x_{1},x_{2})}{\partial x_{2}}} &=&\displaystyle\frac{w_{1}}{w_{2}}%
,  \label{eq:optimality_condition} \\
q(x_{1},x_{2}) &=&\overline{q}.  \label{eq:optimality_condition2}
\end{eqnarray}%
\end{subequations}

The former optimality condition simply states that, in the optimum, the
marginal rate of technical substitution is exactly equal to the ratio of the
factor prices of using $x_{1}$ and $x_{2}$; the latter is just the
output constraint. Note that although the output function is continuous, due to the
interaction between the two inputs it is non-monotonic and defined
differently in different regions of the $(x_{1},x_{2})$ plane -- see Fig. %
\ref{fig:fig1}b. In particular, three regions can be distinguished. Firstly, 
$x_{1}<\kappa /g_1$ and wild bees are present; secondly, $x_{1}\geq \kappa /g_1$
and wild bees are locally extinct. Finally, in a special case when $x_{2}=0$%
, pollination relies exclusively on wild bees. These three regions are
characterised by different optimisation criteria and we first discuss each
one separately. We subsequently compare the results to identify the strategy
associated with the minimum of $c$.

\subsubsection{Regions}

\paragraph{Wild and commercial bees}
For $x_1>0$ and $x_2>0$ the equations (\ref{eq:optimality_condition}) and (%
\ref{eq:optimality_condition2}) can be solved to obtain

\begin{equation}
\begin{array}{r*{20}l} \hat{x}_1 &=& \Delta_1^\alpha\, \bar{q}\\[10pt]
\hat{x}_2 &=& \Delta_1^\alpha\,\displaystyle\frac{\alpha w_1+ w_2
g}{(1-\alpha) w_2} \, \bar{q} - \kappa\\[10pt] \hat{y} &=& \kappa-g_1
\Delta_1^\alpha \bar{q} \, . \end{array}  \label{eq:first1}
\end{equation}%
where $\nu =(1-\alpha )/\alpha $ and%
\begin{equation}
\label{eq:Delta1}
\Delta _{1}=\displaystyle\frac{\nu w_{2}}{w_{1}+w_{2}g}\,.
\end{equation}%
The optimal values of both the pesticide use, $\hat{x}_{1}$, and the
commercial bees use, $\hat{x}_{2}$, are both linear functions of the target
output, $\bar{q}$. Note that $\hat{x}_{2}>0$ if

\begin{equation}  \label{eq:x_2positive}
\bar{q}>\bar{q}_c=\displaystyle\frac{k \Delta_1^{1-\alpha}}{1+g_1 \Delta_1}\, .
\end{equation}

\paragraph{Commercial bees only}
By assuming that $y=\kappa -g_1x_{1}=0$ we can eliminate $x_{2}$ from the
output equation and obtain the cost equation in terms of $x_{1}$ only,%
\begin{equation}
c(x_{1})=w_{1}x_{1}+w_{2}\bar{q}^{1/\alpha }x_{1}^{-\nu }\,.
\label{eq:cost2}
\end{equation}%
Denote%
\begin{equation}
\label{eq:Delta2}
\Delta _{2}=\displaystyle\frac{\nu w_{2}}{w_{1}}\,.
\end{equation}%
Then,%
\begin{equation}
\begin{array}{r*{20}l} \hat{x}_1 &=& \Delta_2^\alpha \bar{q}\\[10pt]
\hat{x}_2 &=& \Delta_2^{1-\alpha}\, \bar{q} \\[10pt] \hat{y} &=& 0 \, .
\end{array}  \label{eq:first2}
\end{equation}%
Again, the optimal values $\hat{x}_{1}$ and $\hat{x}_{2}$ are linear
functions of $\bar{q}$.\footnote{%
Note that, unlike in the previous section, $\hat{x}_{2}>0$ regardless of $%
\bar{q}$.}

\paragraph{No commercial bees} If $x_{2}=0$ then the output equation becomes%
\begin{equation}
q(x_{1},0)=\left( \kappa -g_1x_{1}\right) ^{\alpha }x_{1}^{1-\alpha },
\label{eq:output3}
\end{equation}%
which can be solved to find the corresponding value of $x_{1}$. However, we
can only perform an optimisation if this equation has more than one
solution, otherwise the value of $x_{1}$ is completely determined by the
output level, $\bar{q}$. Unfortunately, no analytical solution can be found
in a general case, but for $\alpha =1/2$ the equation is quadratic and has
two solutions if $\bar{q}<\bar{q}_{m}$, one solution if $\bar{q}=\bar{q}_{m}$%
, and no solutions if $\bar{q}>\bar{q}_{m}$. If equation (\ref{eq:output3})
has two solutions, they are

\begin{equation}
x_{1}^{\pm }=\displaystyle\frac{\kappa }{2g_1}\,\pm \,\displaystyle\frac{\sqrt{%
\kappa ^{2}-4g_1\bar{q}^{2}}}{2g},  \label{eq:x1-3}
\end{equation}%
both of which are positive. As the cost in this case is $%
c(x_{1},0)=w_{1}x_{1}$, the smaller one of these solutions, $x_{1}^{-}$ is
optimal, hence%
\begin{equation}
\begin{array}{r*{20}l} \hat{x}_1 &=&
\displaystyle\frac{\kappa}{2g_1}\,-\,\displaystyle\frac{\sqrt{\kappa^2-4g_1
\bar{q}^2}}{2g_1} \\[10pt] \hat{x}_2 &=& 0 \\[10pt] \hat{y} &=&
\displaystyle\frac{\kappa}{2}\,+\,\displaystyle\frac{\sqrt{\kappa^2-4g_1
\bar{q}^2}}{2} \, . \end{array}  \label{eq:first3}
\end{equation}

In this section we explore how the optimal management options change as the
target output $\bar{q}$ increases. We show that there are three critical
levels of $\bar{q}$ at which the behaviour changes: $\bar{q}_{c}$ is a level
at which commercial bees become economically viable, $\bar{q}_{m}\geq \bar{q}%
_{c}$ is the maximum output achievable without commercial bees, and $%
\bar{q}_{e}\geq \bar{q}_{m}\geq \bar{q}_{c}$ is the output level at which
the optimal use of pesticides leads to local extinction of wild bees. We
subsequently discuss how these threshold levels depend on pesticide and
commercial bee prices, $w_{1}$ and $w_{2},$ and on the carrying capacity, $%
\kappa $. We also discuss potential strategies that a social planner can use
to shift the system from a state in which wild bees are locally extinct to
the state in which they can survive. Finally, we discuss extensions to the
model.

\subsection{Comparative statics}

\begin{figure}[tbph]
\includegraphics[angle=0, width=13cm]{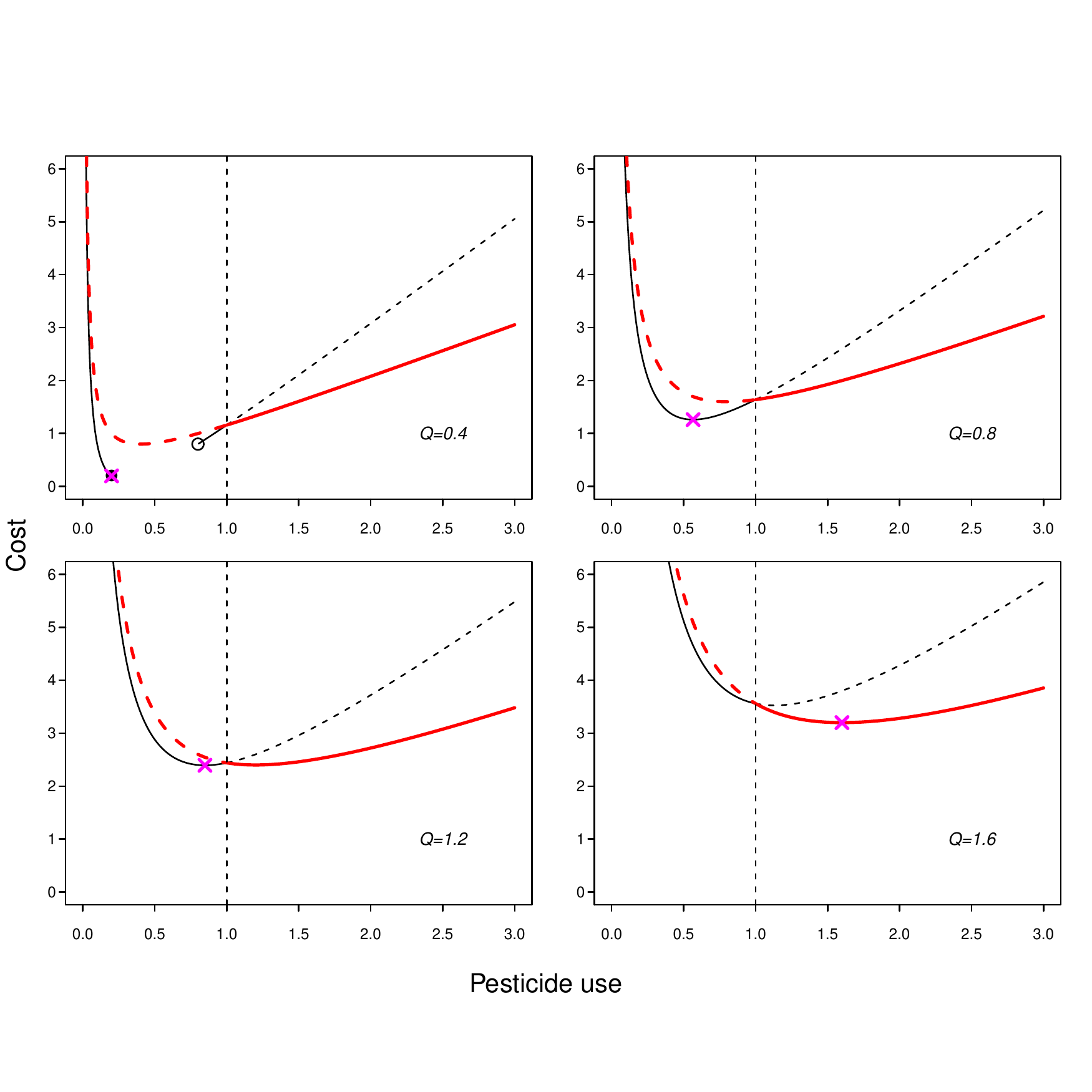}
\caption{Cost, $c\left( x_{1},x_{2}\right) $ as a function of the pesticide
use, $x_{1}$, with $x_{2}$ eliminated through the constraint equation, $%
q(x_{1},x_{2})=\bar{q}$. Graphs correspond to different values of the target
output, $\bar{q}$=0.4 (a), 0.8 (b), 1.2 (c) and 1.6 (d) (see also Fig. 
\protect\ref{fig:fig1}). Thin (black) line represents the case with wild
bees and the thick (red) line represents the case without wild bees (lines
are extended beyond the validity intervals to illustrate the behaviour; the
extensions are marked as broken lines). Vertical line corresponds to $y=%
\protect\kappa -gx_{1}=0$ and wild bees are extinct to the right of it.
Cross represents the location of the optimal solution; solid and empty
circles represent solutions of Eqn. (\protect\ref{eq:output3}), i.e. for $%
x_{2}=0$. Other parameters: $\protect\kappa =1$, $g_1=1$, $w_{1}=1$, $w_{2}=1$
and $\protect\alpha =1/2$.\vskip2cm}
\label{fig:fig2}
\end{figure}

In the previous section we have shown that the optimal strategy $(\hat{x}%
_{1},\hat{x}_{2})$ is different under different assumptions about the values
of $y$ and $x_{2}$. In particular, we have identified three regions: (i) 
\textit{Region 1: }$x_{2}=0$, (ii) \textit{Region 2: } $x_{2}>0$ and $y>0$,
and (iii) \textit{Region 3: } $x_{2}>0$ and $y=0$ --\ see Fig. \ref{fig:fig2}%
. As we, \textit{a priori}, do not know in which of the three regions the
optimal solution would lie for a given value of the target output, $\bar{q}$%
, we first calculate the cost at the optimum, $c(\hat{x}_{1},\hat{x}_{2}),$
using formulas for \textit{all} three regions. As $\hat{x}_{2}$ can be
eliminated using equation (\ref{eq:optimality_condition2}), the value of $%
\hat{x}_{1}$ that corresponds to the lowest cost is selected as an optimal
value; the corresponding pair of the conditional factor demands, $(\hat{x}%
_{1},\hat{x}_{2}),$ describes the optimal management strategy. Let us now
systematically discuss the various interactions in the three different
regions. 

\subsubsection{Region 1: Wild bees only}

We first note that for $x_{2}=0$ the optimisation problem has no real
solutions if $\bar{q}>\bar{q}_{m}=\sqrt{\kappa ^{2}/(2g_1)}$ (we assume here $%
\alpha =1/2$). Thus, $\bar{q}_{m}$ has an interpretation of a maximum target
output that can be achieved by using wild bees only. Note that this
threshold value depends on ecological parameters only and not on any
economic factors. We can now state:

\begin{theorem}
If pollination is provided by wild bees only, there is a maximum output that can be achieved. This output level is determined by ecology of wild bees and their interaction with pesticides.
\end{theorem}

The mechanism for this behaviour is related to the balance between pesticide
use and pollinator population. If the farmer wants to increase the level of
output, she needs to increase the level of pesticide use, which in turn
affects the wild bee population. For small values of $\bar{q},$ and
therefore for small values of $\hat{x}_{1},$ this effect is small and so the
output can be increased. However, for large values of $\hat{x}_{1}$, the
pollinator population is reduced to such an extent that output starts to
decline. Eventually, when $\hat{x}_{1}=k/g_1$, the wild bee population becomes
locally extinct which makes agricultural production impossible (as we assume
that pollination is an essential input and it is performed here by wild bees
only). 
\begin{figure}[tbph]
\includegraphics[angle=0, width=13cm]{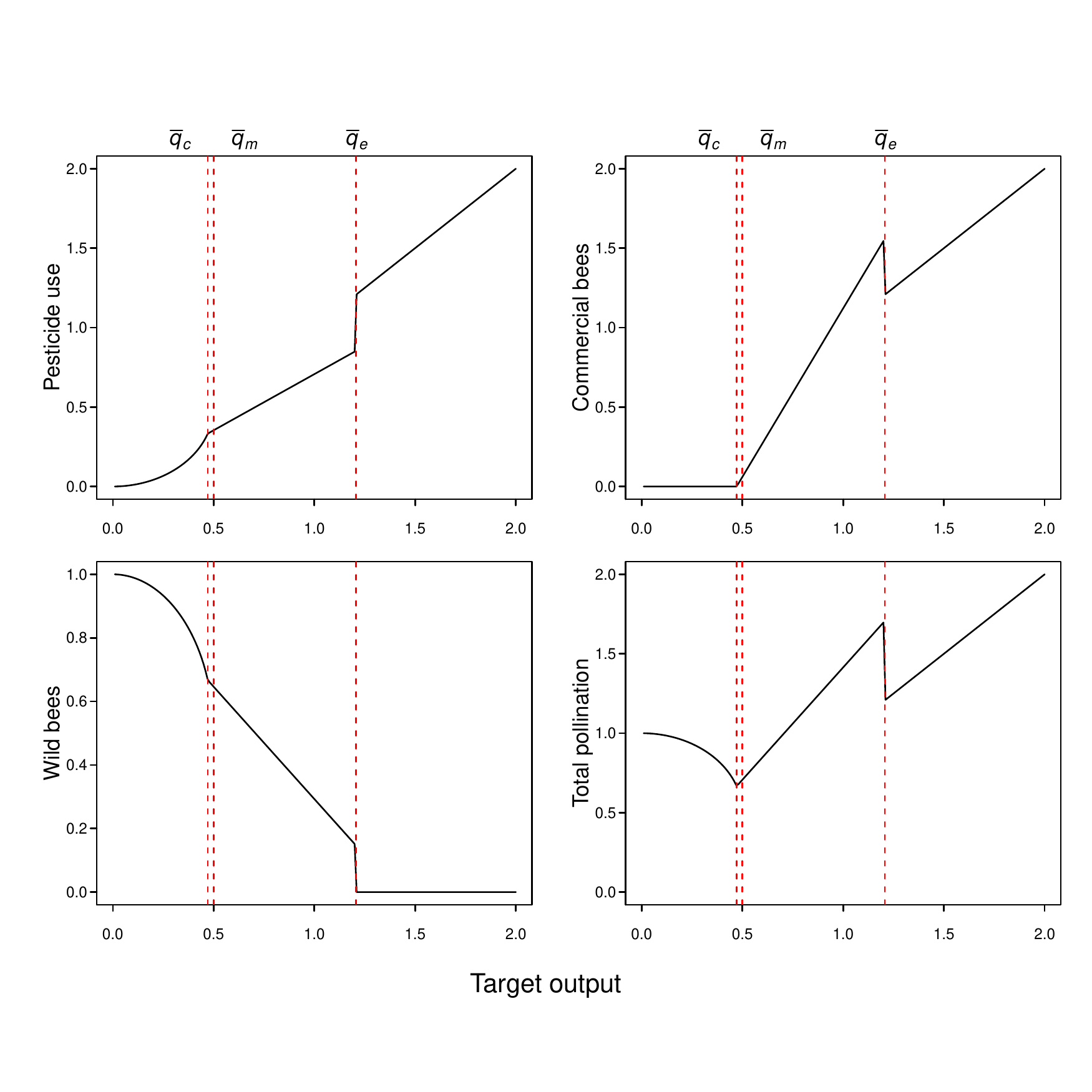}
\caption{Optimum values of (a) pesticide use, $x_{1}$, (b) commercial bees
population, $x_{2}$, (c) wild bees population, $y$, and (d) total
pollination services, $y+x_{2}$, as functions of the target output, $\bar{q}$%
. Vertical lines correspond to the threshold values of $\bar{q}%
_{c}=0.4714045 $, $\bar{q}_{m}=0.5$ and $\bar{q}_{e}=1.207107$,
respectively. Other parameters: $\protect\kappa =1$, $g_1=1$, $w_{1}=1$, $%
w_{2}=1$ and $\protect\alpha =1/2$.\vskip2cm}
\label{fig:fig3}
\end{figure}

\begin{figure}[htbp]
\includegraphics[angle=0, width=13cm]{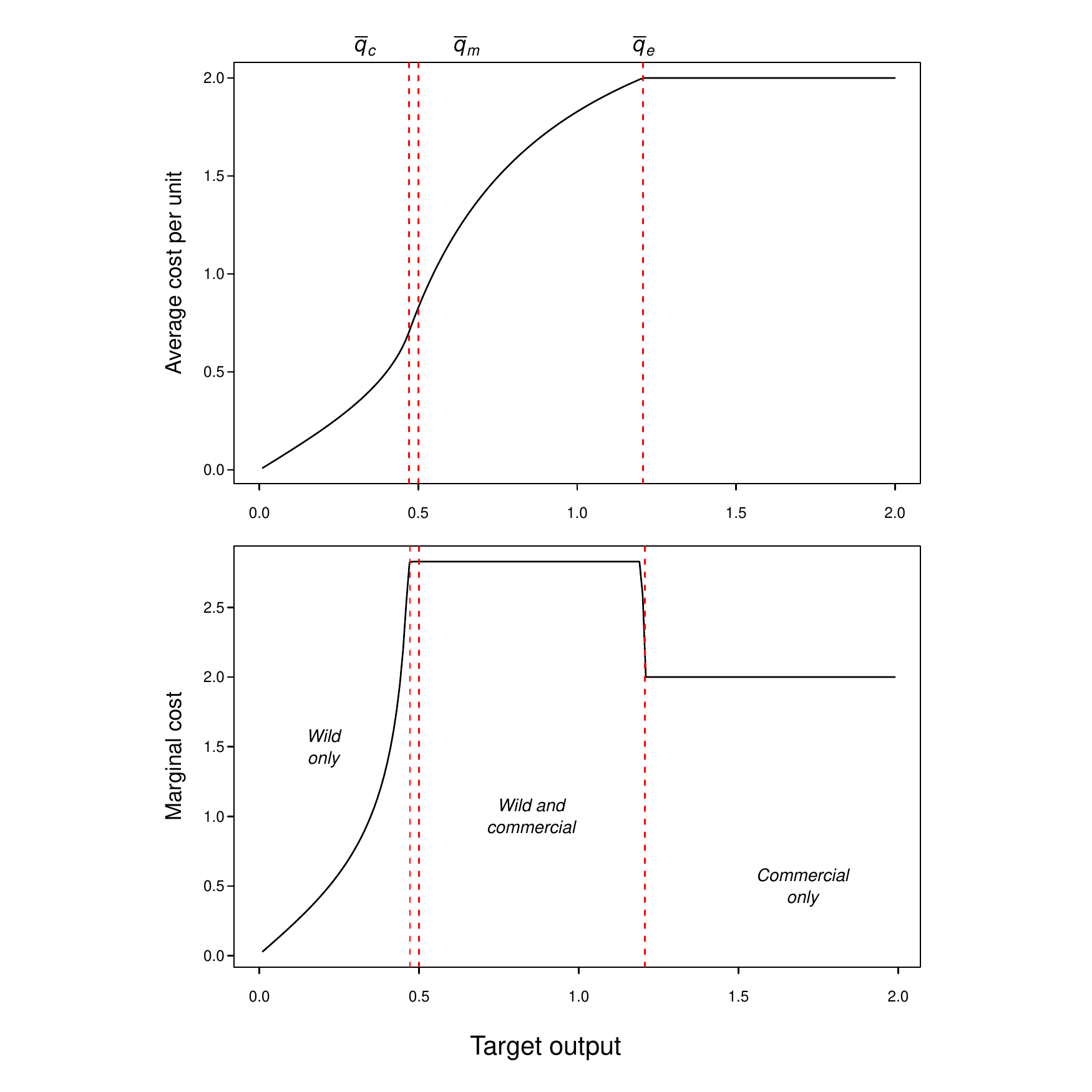}
\caption{The average cost of producing a unit output (a) and the LMC, $%
\partial C/\partial \bar{q}$ (b), as functions of the target output, $\bar{q}
$. Vertical lines correspond to the threshold values of $\bar{q}_c=0.4714045$%
, $\bar{q}_m=0.5$ and $\bar{q}_e=1.207107$, respectively. Other parameters: $%
\protect\kappa=1$, $g_1=1$, $w_1=1$, $w_2=1$ and $\protect\alpha=1/2$.\vskip2cm}
\label{fig:fig4}
\end{figure}

For low values of $\bar{q}$, $x_{1}=x_{1}^{-}$ [\textit{cf}. equation (\ref%
{eq:x1-3})] and $x_{2}=0$ is the optimal choice (see Fig. \ref{fig:fig1}a).
In this case, an increase in the target output, $\bar{q}$, is possible by
increasing the pesticide use (see Fig. \ref{fig:fig3}a), which results in a
gradual decrease in the wild bees population (see Fig. \ref{fig:fig3}c and
the resulting pollination services, Fig. \ref{fig:fig3}d). Equation (\ref%
{eq:shadow3}) below shows the long run marginal cost (LMC), which increases
non-linearly with the target output, $\bar{q}$ (see Fig. \ref{fig:fig4}).
That is, the farmer will find it increasingly more costly to increase the
output by one unit as $\bar{q}$ approaches $\bar{q}_{c}<\bar{q}_{m}$.%
\footnote{%
Note that the LMC becomes infinite at $\bar{q}=\bar{q}_{m}$.}

\begin{equation}  \label{eq:shadow3}
\displaystyle\frac{\partial c}{\partial \bar{q}}=\displaystyle\frac{%
2\delta_1 g \bar{q}}{ g \sqrt{\kappa^2-4g\bar{q}^2}}
\end{equation}

The assumption that $\hat{x}_{2}=0$ is valid as long as $\bar{q}<\bar{q}_{c}$. When $\bar{q}> \bar{q}_{c}$, then  $\hat{x}_2> 0$, see Fig. \ref%
{fig:fig2}b, so there is a sharp transition at $\bar{q}=\bar{q}_c$. As $\bar{q}_{c}<\bar{q}_{m}$, the transition occurs before the
maximum possible output level achievable by using wild bees only is reached
(see Fig. \ref{fig:fig3}). Thus, $\bar{q}_{m}$ is not a good guideline for a
prediction of changes in the behaviour of the combined bioeconomic system.

The results can be summarised as the following proposition:

\begin{theorem}
As the target output level, $\bar{q}$, approaches $\bar{q}_c$, it is increasingly more difficult to increase the output by relying on wild bees only. Thus, commercial bees become an economically attractive option, even though the wild bees  still provide sufficient pollination levels. When $\bar{q}=\bar{q}_c<\bar{q}_m$ is reached, introduction of commercial bees become economically optimal.
\end{theorem}

\subsubsection{Region 2: Wild and commercial bees}

When $\bar{q}>\bar{q}_{c}$,  wild and commercial bees coexist, see Fig. %
\ref{fig:fig2}b and Fig. \ref{fig:fig3}. The pesticide use and the
commercial bee usage increase linearly with $\bar{q}$, while the wild bee
population decreases linearly -- see Fig. \ref{fig:fig3}a and \ref{fig:fig3}%
b, respectively. The LMC in this case does not depend on $\bar{q}$ (see Fig. %
\ref{fig:fig4}):%
\begin{equation}
\displaystyle\frac{\partial c}{\partial \bar{q}}=\Delta _{1}^{\alpha }\,%
\displaystyle\frac{w_{1}+w_{2}g}{(1-\alpha )w_{2}}  \label{eq:shadow1}
\end{equation}%
A desired increase in the target output, $\bar{q}$, is achieved by an
increase in the use of pesticides and the corresponding increase in the
total pollination services (see Fig. \ref{fig:fig3}d). The increase is made
possible by the usage of commercial bees; the total pollination levels
increase, Fig. \ref{fig:fig3}d, although the wild bee population decreases
(see Fig. \ref{fig:fig3}c).

\subsubsection{Region 3: Commercial bees only}

For values of $\bar{q}$ corresponding to Region 2, the optimal costs
calculated by equations (\ref{eq:first1}) and (\ref{eq:first2}) are similar,
but $c(x_{1})$ reaches lower values if $y>0$ (Region 2) -- see Fig. \ref%
{fig:fig2}b. As $\bar{q}$ increases, both curves shift upwards, but at
different rates (\textit{cf}. Fig. \ref{fig:fig2}b with Figs. \ref{fig:fig2}%
c and \ref{fig:fig2}d). At $\bar{q}=\bar{q}_{e}$ the minimum costs using equations (%
\ref{eq:first1}) and (\ref{eq:first2}) are the same. Thus, for $\bar{q}=\bar{%
q}_{e}$ the solution of the optimality problem is not unique; selection of
two different combinations of $(\hat{x}_{1},\hat{x}_{2})$ (both of which
satisfy $q(\hat{x}_{1},\hat{x}_{2})=\bar{q}$) results in identical cost
values. This transition can be also seen in Fig. \ref{fig:fig1}, where the
straight line corresponding to the cost function touches the isoquant line
at two places. The threshold value can be found by equating the optimal
costs calculated from (\ref{eq:first1}) and (\ref{eq:first2}):%
\begin{equation}
\bar{q}_{e}=\,\displaystyle\frac{\kappa w_{2}^{2}(1-\alpha )}{%
(w_{1}+w_{2}g)\Delta _{1}^{\alpha }-(1-\alpha )w_{2}\left( w_{1}\Delta
_{2}^{\alpha }+w_{2}\Delta _{2}^{1-\alpha }\right) }\,.  \label{eq:q0}
\end{equation}

For $\bar{q}>\bar{q}_{e},$ the optimum given by (\ref{eq:first2}) is lower.
This solution corresponds to $\hat{x}_{1}>\kappa /g_1$ and so to $y=0$,
resulting in the local extinction of wild bees. To understand this
transition, we need to look at output levels which depend on both pesticide
use and pollination services. At $\bar{q}=\bar{q}_{e}$ the management
strategy undergoes a substantial shift. Instead of relying largely on the
increase in the pollination services (see Fig. \ref{fig:fig3}d), the farmer
switches to a high use of pesticides (see Fig. \ref{fig:fig2}a) and a lower
use of commercial bees (see Fig. \ref{fig:fig2}b). Note that the further
increase in the target output will be achieved by increasing the pesticide
use more than by increasing the commercial bee population; compare the
slopes in Figs. \ref{fig:fig3}a and \ref{fig:fig3}b for $\bar{q}<\bar{q}_{e}$
and for $\bar{q}>\bar{q}_{e}$.\footnote{%
This can also be shown by noting that $\Delta _{1}<\Delta _{2},$ hence the
slope of $x_{1}$ as a function of $\bar{q}$ is larger in Region 3 than in
Region 2.}

The shift in the management strategy causes the wild bees population to
collapse (see Fig. \ref{fig:fig2}c). The total pollination services go down
at $\bar{q}=\bar{q}_{e}$, but continue to increase afterwards. Again, the
LMC does not depend on $\bar{q}$ and for the parameters used here is lower
than in Region 2:%
\begin{equation}
\displaystyle\frac{\partial c}{\partial \bar{q}}=w_{1}\Delta _{2}^{\alpha
}+w_{2}\Delta _{2}^{1-\alpha }.  \label{eq:shadow2}
\end{equation}%
This adds another incentive for farmers to switch to the new management
strategy --\ leading to extinction of wild bees -- as this will not only
allow to increase their output, but also to lower its marginal costs.

\begin{theorem}
A high target output and a lower LMC can be achieved by increasing the reliance on pesticide use rather than pollination, leading to the local extinction of wild bees.
\end{theorem}

\subsection{Extending the model}

In this section we return to the general model and discuss how relaxing the simplifying assumptions affects the results. In particular, we consider three extensions: the differential pollination potential of wild and commercial bees (assumption (iii)), the influence of pesticides on commercial bee population (assumption (iv)), and the asymptotic behaviour of the production function with respect to $x_1$ and $x_2$ (assumption (ii)). Relaxation of assumptions (i) and (v) exceeds the scope of this paper.

\subsubsection{Differential pollination}

First, we analyse the effect of $u\not =1$, i.e. different ability of wild and commercial bees to pollinate the crop, on the behaviour of the simplified model. In this case, the total pollination potential depends on $u x_2$ instead of $x_2$. Defining $\bar{x}_2=u x_2$ and considering $\bar{x}_2$ as the control variable leads to analogous results as for the simplified model, except that the cost function is now

\begin{equation}
w_{1}x_{1}+w_{2}x_{2}=w_{1}x_{1}+\dfrac{w_{2}}{u}\bar{x}_{2}\,, \label{eq:costbase-rho}
\end{equation}%

i.e. corresponds to a rescaled price for commercial bees, $w_2$. All the results from above apply with the appropriate scaling of $w_2$. Thus,

\begin{theorem}
If commercial bees are more efficient in pollinating the particular crop than wild bees, $u\gg 1$, they are more likely to be introduced (lower $\bar{q}_c$), if the unit price is the same. 
\end{theorem}

\subsubsection{Effect of pesticides on commercial bees}

In the analysis above we assumed that $g_2=0$ so the ability of commercial bees to pollinate crop is not affected by pesticides. If $g_2\not =0$ then $z=x_2-g_2 x_1$ if $x_2-g_2 x_1>0$ and 0 otherwise. We will assume that the effect on commercial bees is smaller than on wild bees so that $g_2\ll g_1$. Equations (\ref{eq:bees}) and (\ref{eq:costbase}) can be simplified by expressing the costs in terms of $z$ rather than $x_2$,

\begin{equation}
w_{1}x_{1}+w_{2}x_{2}=w_{1}x_{1}+w_{2}(z+g_2 x_1)=(w_{1}+g_2w_{2})x_{1}+w_{2}z\,. \label{eq:costbase-compest}
\end{equation}%

Assume first that $\kappa-g_1 x_1> 0$ and $x_2-g_2 x_1>0$, i.e. the pesticide use, $x_1$, is small. Then,

\begin{equation}
q\left( {x_{1},z}\right) ={\left( {\kappa -g_1 x_{1}+z}\right)
^{\alpha }}{x_{1}^{1-\alpha }}= \bar{q}\,.  \label{eq:profit1-compest}
\end{equation}%

This is the same problem as our simplified model, but with $z$ being the control variable instead of $x_2$, and with the price for pesticides modified by the addition if the $g_2 w_2$ term. This has a simple interpretation: the farmer aims to control the effective population of commercial bees by appropriately increasing the purchased stock $x_2$, with the increase given by $g_2 x_1$. This results in an increased effective price for pesticides, as their effect on commercial bees needs to be taken into account.

If the pesticide use is increased so that $\kappa-g_1 x_1\leq 0$ but $x_2-g_2 x_1>0$ (assuming here $g_2\ll g_1$), then

\begin{equation}
q\left( {x_{1},z}\right) ={z^{\alpha }}{x_{1}^{1-\alpha }}= \bar{%
q}\,.  \label{eq:profit1for0-compest}
\end{equation}%

This is again the same problem as for our simplified model. Proceeding as above, we obtain

\begin{equation}
\label{eq:Delta2-compest}
\Delta _{2}=\displaystyle\frac{\nu w_{2}}{w_{1}+w_2 g_2}\,.
\end{equation}%
Then,%
\begin{equation}
\begin{array}{r*{20}l} \hat{x}_1 &=& \Delta_2^\alpha \bar{q}\\[10pt]
\hat{z} &=& \Delta_2^{1-\alpha}\, \bar{q} \\[10pt]
\hat{x_2} &=& \left(\Delta_2^{1-\alpha}+g_2 \Delta_2^\alpha\right) \bar{q}= \\[10pt]
 \hat{y} &=& 0 \, ,
\end{array}  \label{eq:first2-compest}
\end{equation}%

which proves that $\hat{z}$ will always be positive (i.e. the use of pesticides and the input of commercial bees will always be adjusted so that pollination is supplied). The total cost, however,

\begin{equation}
\label{eq:Cost-compest}
\left(w_1\Delta_2^\alpha + w_2\Delta_2^{1-\alpha}+w_2 g_2 \Delta_2^\alpha\right) \bar{q} \,.
\end{equation}%

is now increased by the cost of purchasing additional commercial bees, $w_2 g_2 \Delta_2^\alpha$, as compared to the case when $g_2=0$. Thus,

\begin{theorem}
If commercial bees are affected by pesticides,  behaviour is qualitatively the same as above, but there is an associated cost of purchasing additional bees to compensate this effect. This can be described by an increase in an effective unit price of pesticides.
\end{theorem}

\subsubsection{Asymptotic behaviour}

\begin{figure}[htbp]
\includegraphics[angle=0, width=13cm]{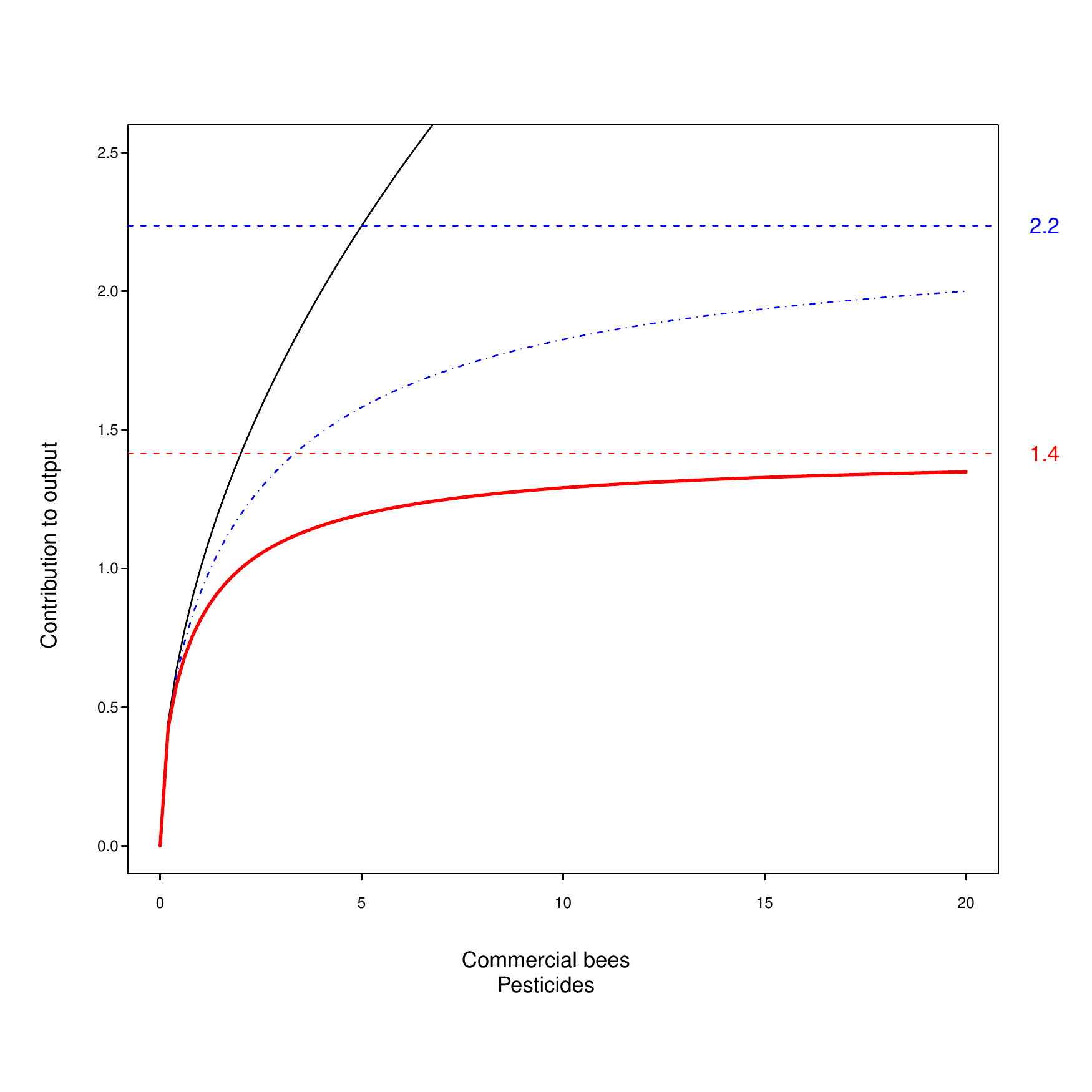}
\caption{Functional forms for the output dependence on pollination (here on
commercial bees only), for the standard Cobb-Douglas function, Eqn. (\protect\ref%
{eq:profitbase}) with $\protect\eta=0$ (thin line), and for the nonlinear
form with $\protect\eta=0.1$ (dotted line), and with $\eta=0.5$ (thick line).
The graphs are the same for pesticide use ($\protect\rho=0$, $\protect\rho%
=0.1$, $\protect\rho=0.5$, respectively). Horizontal lines show the
respective asymptotes when applicable, also indicated in the margin. $%
\protect\alpha=\beta=1/2$.\vskip2cm}
\label{fig:newfig}
\end{figure}

In the simplified model considered above, equations (\ref{eq:profit1})-(\ref{eq:problem}), farm output is an increasing function of the
pollinator population size (both wild and commercial), but without an asymptote (see Fig. \ref{fig:newfig}).
Thus, by increasing the commercial bee density one can make output to be
arbitrarily large. In reality, the functional relationship between
pollinator population size and the delivery of pollination services is
likely to be an asymptotic type of function, so that output cannot increase
unlimitedly. Similar features characterises
pesticide use. Adding nonlinear functional forms for $y$, $x_{1}$ and $x_{2}$
as in equation (\ref{eq:production}) significantly complicates the model and, therefore, we only consider Region
2 which corresponds to a high level of pesticide use and a high commercial
bees density (and to $y=0$). This is the range where the saturation effect
is most likely to occur; this assumption will be relaxed in future
work. For simplicity we also assume $\alpha =\beta=1/2$.

Given this setting, let us consider the minimization of the cost $%
c(x_{1},x_{2})=w_{1}x_{1}+w_{2}x_{2}$ subject to the following constraint:

\begin{equation}
q(x_{1},x_{2})=\left( B(x_{2})\right) ^{1/2}\left(P(x_{1})\right)^{1/2}=\bar{q},
\label{eq:asymptotic_output}
\end{equation}%
where it is assumed that $y=0$. Firstly, assume that the pesticide
functional form is linear, $P(x_{1})=x_{1}$ (the Cobb-Douglas form still applies so that second derivative of $q$ with respect to $x_1$ is negative), but the pollination production function tends asymptotically to a limit according to:%
\begin{equation}
B(x_{2})=\displaystyle\frac{x_{2}}{1+\eta x_{2}},  \label{eq:highx2form}
\end{equation}%
where $1/\eta $ is the asymptotic value for the pollination service (see
Fig. \ref{fig:newfig}). The optimal values read as:%
\begin{equation}
\begin{array}{r*{20}l} \hat{x}_1&=&\sqrt{\dfrac{w_2}{w_1}}\,\bar{q}+\eta
\bar{q}^2\\[10pt] \hat{x}_2&=&\sqrt{\dfrac{w_1}{w_2}}\,\bar{q}\\[10pt]
c(\hat{x}_1,\hat{x}_2)&=&2\sqrt{w_1 w_2}\,\bar{q}+\eta w_1 \bar{q}^2 \, .
\end{array}  \label{eq:highx2results}
\end{equation}%

Note that given $\alpha =1/2$, $v$ becomes equal to 1 and $\Delta _{2}$ is simplified to $w_{2}/w_{1}$ (\textit{cf}. Eqn. (\ref{eq:Delta2})) so the
above result is compatible with equation (\ref{eq:first2}) if $\eta =0$.
Thus, $\hat{x}_2$ is unaffected by the nonlinearity in the pollination production function. However, the pesticide use, $\hat{x}_1$ must increase to offset the relative inefficiency
of pollination. Consequently, the LMC now increases as a function of the target output $\bar{q}$:%
\begin{equation}
\displaystyle\frac{\partial c}{\partial \bar{q}}=2\sqrt{w_{1}w_{2}}+2\eta
w_{1}\bar{q}\,.  \label{eq:highx2shadow}
\end{equation}

Secondly, assume that the pollination production function is not limited, so $B(x_{2})=x_{2}$  (the Cobb-Douglas form still applies so that second derivative of $q$ with respect to $x_2$ is negative), but that the pesticide efficiency is an
asymptotic function of $x_{1}$:

\begin{equation}
P(x_{1})=\displaystyle\frac{x_{1}}{1+\rho x_{1}},  \label{eq:highx1form}
\end{equation}%
where $1/\rho $ is the asymptotic value for the pesticide efficiency (see
Fig. \ref{fig:newfig}). The optimal values are then:%
\begin{equation}
\begin{array}{r*{20}l} \hat{x}_1&=&\sqrt{\dfrac{w_2}{w_1}}\,\bar{q}\\[10pt]
\hat{x}_2&=&\sqrt{\dfrac{w_1}{w_2}}\,\bar{q}+\rho \bar{q}^2\\[10pt]
c(\hat{x}_1,\hat{x}_2)&=&2\sqrt{w_1 w_2}\bar{q}+\rho w_2 \bar{q}^2 \, .
\end{array}  \label{eq:highx1results}
\end{equation}%
As before, equation (\ref{eq:highx1results}) is compatible with equation (%
\ref{eq:first2}) if $\rho =0$. The solution mirrors the previous case in
that the optimal use of pesticides is unaffected but that the commercial bee
population must increase to offset the inefficiency in pest control. The
cost, again, increases and the additional term is a quadratic function of $%
\bar{q}$; the LMC increases with $\bar{q}$:

\begin{equation}  \label{eq:highx1shadow}
\displaystyle\frac{\partial c}{\partial \bar{q}}=2\sqrt{w_1 w_2}+2\rho w_2 
\bar{q} \, .
\end{equation}

Finally, if both pollination and pesticide production functions have asymptotes, i.e. $\eta >0$ and $\rho>0$, then:

\begin{equation}
\begin{array}{r*{20}l} \hat{x}_1&=&\displaystyle\frac{\bar{q}}{1-\eta\rho
\bar{q}^2}\left(\sqrt{\dfrac{w_2}{w_1}}+\eta \bar{q} \right)\\[10pt]
\hat{x}_2&=&\displaystyle\frac{\bar{q}}{1-\eta\rho
\bar{q}^2}\left(\sqrt{\dfrac{w_1}{w_2}}+\rho \bar{q}\right)\\[10pt]
c(\hat{x}_1,\hat{x}_2)&=&\displaystyle\frac{\bar{q}}{1-\eta\rho
\bar{q}^2}\left(2\sqrt{w_1w_2}+(w_1\eta+w_2\rho) \bar{q} \right)\, .
\end{array}  \label{eq:highx1x2results}
\end{equation}%
In this case it is not possible to satisfy an arbitrarily increasing
target output and the maximum possible $\bar{q}$ is given by $1/\sqrt{\eta
\rho }$. This is due to the fact that one cannot offset a saturation in one
factor (say, $x_{1}$) by increasing the other (say, $x_{2}$). Both $x_{1}$
and $x_{2}$ need to be set very high, resulting in a disproportional cost increase.\footnote{%
We do not explicitly state the LMC in this case, but it tends to infinity as 
$\bar{q}\rightarrow 1/\sqrt{\eta \rho }$.} This leads us to contend:

\begin{theorem}
Introduction of saturating functional forms for $x_1$ or $x_2$ does not change the results qualitatively for Region 2, but increases the cost and makes LMC a nonlinear function of $\bar{q}$. If the contribution of both pesticide use and pollination services is of an asymptotic form, there is a maximum output that can be achieved.
\end{theorem}

We cannot obtain the thresholds $\bar{q}_{c}$ and $\bar{q}_{e}$ analytically
in this case, but one can draw some general conclusions by observing that
higher levels of $x_{1}$ (and $x_{2}$) are needed to produce the same output
level, if $\rho >0$ ($\eta >0$\/). This will cause wild bees to become
extinct at lower levels of the target output, $\bar{q}$, as compared to the
standard Cobb-Douglas equation. We therefore expect the threshold values $%
\bar{q}_{c}$ and $\bar{q}_{e}$ to be lower in this case.

\section{Discussion}

\label{sec:discussion}

In the paper we studied the dependence of cost-minimising management
strategies on the target farm output, $\bar{q}$. Thus, given the output target (for example set in the contract specifying delivery to a supermarket chain), the
farmer will chose a certain strategy, minimising the private costs of
output. We have shown that depending on the $\bar{q}$, different strategies
emerge, in which it is more profitable for the farmer to either refrain from
using commercial bees when $\bar{q}<\bar{q}_{c}$, or to use them if $\bar{q}%
\geq \bar{q}_{c}$. Use of commercial bees allows the farmer to move beyond
yield levels that can be achieved by natural pollination ($\bar{q}>\bar{q}%
_{m}$\/). This is achieved by increasing both pesticide use (Fig. \ref%
{fig:fig3}a) and the total pollination levels (Fig. \ref{fig:fig3}d).

Interestingly, we also show that there are two competing strategies that the
farmers can use to achieve an output exceeding $\bar{q}_{e}$: a
low-pesticide, high-pollination strategy or a high-pesticide,
low-pollination strategy (\textit{cf}. Fig. \ref{fig:fig3}a and Fig. \ref%
{fig:fig3}d). There are good reasons to move to the latter management
strategy, as it leads to a lower long run average cost (LAC) of producing a
unit output (see Fig. \ref{fig:fig4}a) and a lower long run marginal cost
(LMC) if $\bar{q}\geq \bar{q}_{e}$ (see Fig. \ref{fig:fig4}b). However, this
strategy also leads to local extinction of wild bees if $\bar{q}\geq \bar{q}%
_{e}$ holds.

Local extinction occurs because the link between the wild bee population and
output is broken, and the farmer does not get a signal that wild
bees are declining. This is caused by the availability of commercial bees.
However, we show that the ecological effect of introduction of commercial
bees depends on the intensity of production. In the low-intensity production
situation, the introduction of commercial bees leads to an increase in the
use of pesticides and therefore to a decrease in the wild bees populations
(see Fig. \ref{fig:fig3}). 

We also note that the transition at $\bar{q}=\bar{q}_{e}$ is an abrupt one.
The wild bees population might be relatively low but healthy (ca. 20\% of
the carrying capacity for parameters used in this paper -- Fig. \ref%
{fig:fig3}c) for $\bar{q}$ smaller but close to $\bar{q}_{e}$. However, for $%
\bar{q}>\bar{q}_{e}$, farmers are likely to switch to a high-pesticide
strategy and the population of wild bees will then likely become extinct.
Thus, a small change in the target output, caused for example by a surge in
soft fruit prices, can make the high-pesticide use strategy economically
attractive, leading to a dramatic decline in the wild bees population.

The threshold output $\bar{q}_{e}$ is therefore very important from a policy
perspective. The social planner faced with a system in which wild bees are locally extinct, might want to formulate policies leading to their re-establishment. The simplest policy is to encourage farmers to lower their target output, $\bar{q}$ below the threshold value, $\bar{q}_e$. Alternatively, the social planner might want to increase the threshold value,  $\bar{q}_e$. This can be achieved by: increasing the price of pesticides (increasing $w_{1}$\/)\footnote{%
This could be done by imposing a tax on pesticide use, for instance.},
encouraging farmers to stimulate wild bee population (increasing $\kappa $%
\/), switching to alternative pesticides (decreasing $g$\/), and/or
decreasing the price of commercial bees (decreasing $w_{2}$\/). Although the
formula for $\bar{q}_{e}$ is known [see equation (\ref{eq:q0})], and it
depends on both ecological ($\kappa $, $g$\/) and economic ($w_{1}$, $w_{2}$%
, $\alpha $\/) factors, it is not clear whether it can reliably be estimated
in practice. This means that the threshold might be difficult to predict in
advance and the transition between management strategies might be difficult
to incentivise \citep{Taylor2009}.

\begin{figure}[htbp]
\includegraphics[angle=0, width=13cm]{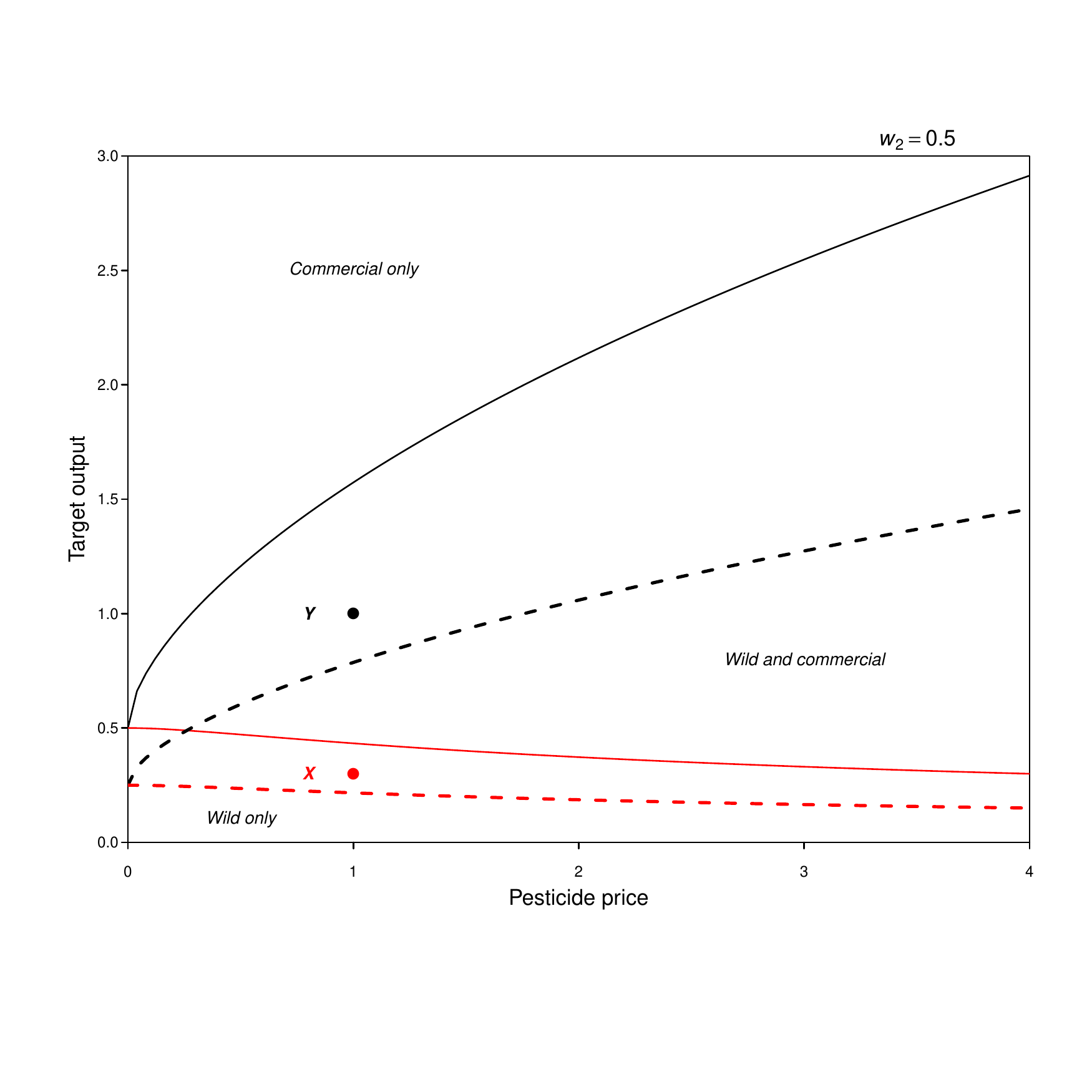}
\caption{The effect of changing $\protect\kappa$ on the ecological outcome
under the optimal management strategy as a function of the marginal
pesticide cost, $w_1$ and the target output, $\bar{q}$; $\protect\kappa=1$
(solid lines) is reduced to $\protect\kappa=0.5$ (broken lines). Lines
represent the threshold budget values, $\bar{q}_c$ (black line) and $\bar{q}%
_e$ (red line). Other parameters: $g_1=1$ and $\protect\alpha=1/2$; $w_2=0.5$. 
$X$ and $Y$ strategies are discussed in the text.\vskip2cm}
\label{fig:fig7}
\end{figure}

While we have focused on pesticide use, the switch can also be caused by
other factors. For example, if the carrying capacity of wild bees, $\kappa $%
, decreases due to a reduction in the size or quality of wild bee-friendly habitat, the
threshold values $\bar{q}_{c}$ and $\bar{q}_{e}$ will decrease (see
equations (\ref{eq:x_2positive}) and (\ref{eq:q0})). If the farmer still
wants to attain the same target output, she will need to change the
strategy, depending on the combination of the target output and the
pesticide and commercial bees prices. If the target output is low, under
current conditions ($\kappa =1$), the farmer does not need to use commercial
bees (see strategy X in Fig. \ref{fig:fig7}).

A decrease in the carrying capacity, for example triggered by bad weather
conditions or changes in land use causing a reduction in foraging or nesting
areas for wild bees, leads to a shift in the optimal farm strategy (see Fig. \ref{fig:fig7}) -- point X now lies in Region 2
demarcated by the broken lines. Thus, the farmer might feel that the change in environmental conditions forces her to introduce commercial bees. 

If the level of output is high and the
farmer already uses commercial bees -- strategy Y in Fig. \ref{fig:fig7} --
the reduction in the carrying capacity means switching to a
high-pesticide use (as point Y now lies in Region 3) with an associated
local extinction in the wild bee population.

As noted above, the change in the wild bee population as the system moves
from Region 2 to Region 3 is very abrupt. For instance, we might be dealing
with what looks like a farm with a healthy population of wild bees (although
smaller than their carrying capacity) for one set of environmental conditions $%
\kappa $. A small change in the environmental conditions results in a
limited loss in their ability to pollinate which in turns leads to the
farmer modifying her management practices to keep up with demand for
agricultural produce. This modification results in a rapid local extinction
of wild bees as the threshold, $\bar{q}_{e}$ is crossed. The importance of
maintenance of a conducive environment for wild bees has also been
emphasised by \citet{Keitt2009}, who  concluded that bee populations
would decline abruptly if environmental conditions (in this case, habitat)
deteriorated below a certain threshold. The results of this paper support
this conclusion, as increasing the carrying capacity of the surrounding area
would allow higher outputs without exceeding the threshold leading to
population extinction.

The particular functional forms are chosen in this paper for their general
applicability to a wide range of agri-ecological problems as well as for
their simplicity; this applies to both the linear dependence of $B$ on $x_1$ and to the Cobb-Douglas production function.
 The latter has been chosen over alternative forms to model pollination services
because it reflects a complete reliance of the production system on wild or
commercial bees. This assumption will be accurate for many crops with a high
dependence on pollinators, such as many berries and orchard fruits \citep{Hanley2015}. In
particular, if $y$ and $z$ are both zero, the output is zero,
independently of $x_{1}$. Although in our paper $x_{1}$ represents the use
of pesticides, it can also be interpreted as any agricultural practice that
(i) is essential for the output generation, and (ii) affects wild bee
populations. This generalisation strengthens the case for the
particular functional form. Finally, we assumed that commercial bees do not directly compete with wild bees. This assumption can be relaxed and we expect this to make the population of wild bees more fragile. This extension would require a more realistic population model and exceeds the scope of this paper. 

In our model the changes are fully reversible (the wild bees population
reacts immediately to changes in $x_{1}$), but in reality such shifts are
likely to be irreversible. This will be problematic if the strategy taken
then becomes uneconomic, for example due to an increase in the price of
pesticides, or commercial bees. Such a switch has already occurred in the
apple growing region of Sichuan, China, where human pollinators were used as
substitutes, allowing a high pesticide, low habitat strategy to continue.
When human pollination became too expensive, the only option for farmers was
to leave the market altogether, and discontinue apple production. When
declines in wild capital such as wild pollinators are irreversible, and
there is uncertainty over which source of capital will be most beneficial in
the future, there is a value to maintaining the natural capital for future
use \citep{Arrow1974,Kassar2004}. This ``option''
value is an incentive for conserving wild pollinators, and will be positive
even if there are no immediate advantages of supporting wild pollinators.

The wild bee population modelled here will often be made up of multiple
populations of bee and non-bee pollinators (such as hover flies). The
presence of multiple pollinator groups could buffer the system to
extinction; the relative tolerance of pollinator networks to extinction has
been shown by \citet{Memmott2004,Kaiser-Bunbury2010}. However these studies
do not assume that threats to the different populations are correlated.
While different pollinators groups may respond in slightly different ways to
external pressure such as pesticide use, the effects are likely to be
negative on all groups, and may be stronger on non-bee pollinators as these
are smaller \citep{Goulson2013}. The model discussed in this paper is unique
in its inclusion of a chronic threat to pollinators (pesticide use), which
is likely to affect all pollinator groups. The benefit of maintaining
multiple groups of ecosystem service providers as insurance against a
fluctuating environmental was discussed by \citet{Baumgartner2007}. The
problem considered here differs as we consider a threat which is likely to
be detrimental on the whole pollinator community, means that holding diverse
pollinators will not be beneficial, however maintaining both commercial and
wild bees will be valuable as options for pollination provision in the
future.

In our model farmers act myopically. We also assume that wild bees respond instantaneously to the changes
in the management. In effect, the model demonstrates behaviour that would be
observed if the farmer can make planning decisions once and see the impact
of those in the future without the chance of adaptation. This may well be
realistic, as farmers are unlikely to be able to detect small changes in
wild bee populations from year to year, but will notice dramatic decreases
in pollination services over longer periods of time. However, the model, and
in particular its agent-based extension, can be generalised to include
different planning horizons for farmers  as well as the long-term dynamics of bees, for example in the form of
a dynamic model of \citet{Khoury2011}.

\textbf{Acknowledgements:} We are grateful to the European Investment Bank (EIB) University Research Action Programme for financial support of this work through the ECO-DELIVERY project. Any errors remain those of the authors. The findings, interpretations and conclusions presented in this article are entirely those of the authors and should not be attributed in any manner to the EIB.

\bibliographystyle{rspublicnat}
\bibliography{bioecon-AKCE1.bib}

\end{document}